\documentclass[showpacs,amsmath,amssymb,twocolumn]{revtex4}

\usepackage{graphicx}
\input{epsf}
\newcommand{\ro}{ \mbox{\boldmath{$\rho$}}}
\begin{document}

\title{Photogalvanic current in a double quantum well}

\author{M.V. Entin$^{1}$, L.I. Magarill$^{1,2}$}
\affiliation{$^{1}$Institute of Semiconductor Physics, Siberian
Branch of Russian Academy of Sciences, Novosibirsk, 630090, Russia
\\ $^{2}$Novosibirsk State University, Novosibirsk, 630090, Russia}
\begin{abstract}
We study the in-plane stationary current caused by
phototransitions between the states of a double quantum well. The
electric polarization of light has both vertical and in-plane
components. The stationary current originates from the periodic
vibration of electrons between two non-equivalent quantum wells
caused by the normal component of the alternating electric field
with simulteneous in-plane acceleration/deceleration by the
in-plane component of electric field. The quantum mechanism of the
stationary current is conditioned by in-plane transition asymmetry
which appears due to the indirect phototransitions with the
participation of impurity scattering.  The photocurrent has a
resonant character corresponding to the equality of the photon
energy to the distance between subbands. It is found that the
current appears as a response to the linear-polarized light.
\end{abstract}

   \maketitle

\section{Introduction}
 Since the first studies on the photogalvanic effect (PGE) at the end of the 70th, a wide  literature devoted to this subject has appeared \cite{malin,we,we1}, see
 also reviews \cite{belin,ivch,sturm,ivch1,ivch2,gan}. The activity in this field continues up
  to now (see, for example, \cite{shepel,shepel1,we2,shepel2,sassin,ivch3,taras,we3,karch}). There are different variants of
  PGE  in confined systems:  the stationary  in-plane photocurrent in classical  \cite{we4} and quantum \cite{we5,taras1} films,  and the current along solid-state surface\cite{alper,gus,we6}.
  This photocurrent exists even if crystal asymmetry is negligible,
  but the quantum well is oriented (directions across the well are not equivalent).
  The current along the surface occurs if the electric field of the light has both in- and
  out-plane components.

The phenomenology of PGE  is determined by the relation for the current density
\begin{equation}\label{phenomen}
   {\bf j}=\alpha_s(( {\bf E}-{\bf n}(\bf nE))({\bf n
   E^*}+c.c)+i\alpha_a[{\bf n}[{\bf E}{\bf E^*}]],
\end{equation}
where ${\bf n}$ is the normal to the quantum well. Real constants
$\alpha_s$ and $\alpha_a$ describe linear and circular
photogalvanic effects, correspondingly. The origin of this current
can be understood if to consider the out-of-plane component of
electric field as modulating the quantum well conductivity with a
simultaneous driving of electrons by the field in-plane .

In a quantum well the vertical component of the electric field of
light can cause the transitions  between different quantum
subbands. In the presence of scattering this gives birth to the
effective pumping of the in-plane momentum to the electronic
system. The light plays the role of the energy and
non-equilibrium source, while the scatterers produce electrons in-plane
acceleration. The situation is, in a certain  sense,
similar to the motion of a car where the friction forces the car
to move.

The purpose of the present article is to study the mechanism of
PGE in a double quantum well. This system looks
perspective because  the structure of the levels of a double quantum
well permits easy tuning of the distance between subbands to the
frequency of the external field.

The effect under consideration is illustrated in Fig.1.  We
consider intersubband transitions of electrons in a system with the
quadratic energy spectrum. An electron goes between two states
$\epsilon_n({\bf p})$ and $\epsilon_{n'}({\bf p}')$ due to the
simultaneous action of electric field and scattering. These states
originate from mixing  the states of different individual
quantum wells. The in-plane current appears  due to the change of
electron in-plane momentum. To ''memorize'' electric field in- and out-plane
components,  the transition probability should
contain their product. For non-conservation of the electron
momentum the scattering should be taken into account.  This
transition probability arises in the second order of the
perturbation theory. The amplitude of transitions has a resonance
on an intermediate state. The subbands of the quantum well are
equidistant, that gives rise to the absence of the resonance smearing
due to the difference in electron momenta.  The result of excitation
is the pumping of the momentum to the electron subsystem and the
in-plane current.
\begin{figure}[h]\label{fig1}
\centerline{\epsfxsize=6cm\epsfbox{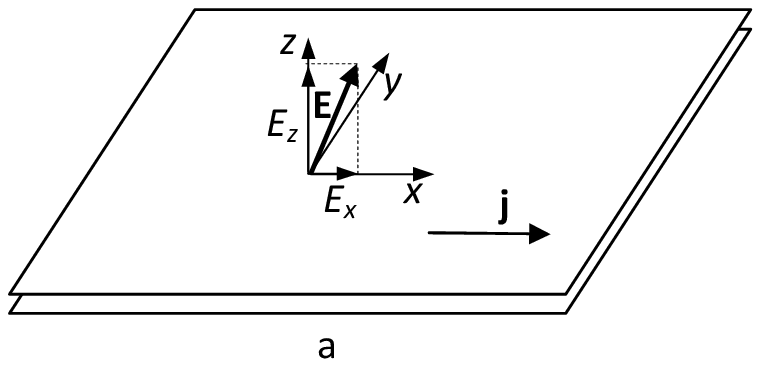}}\centerline{\epsfxsize=6cm\epsfbox{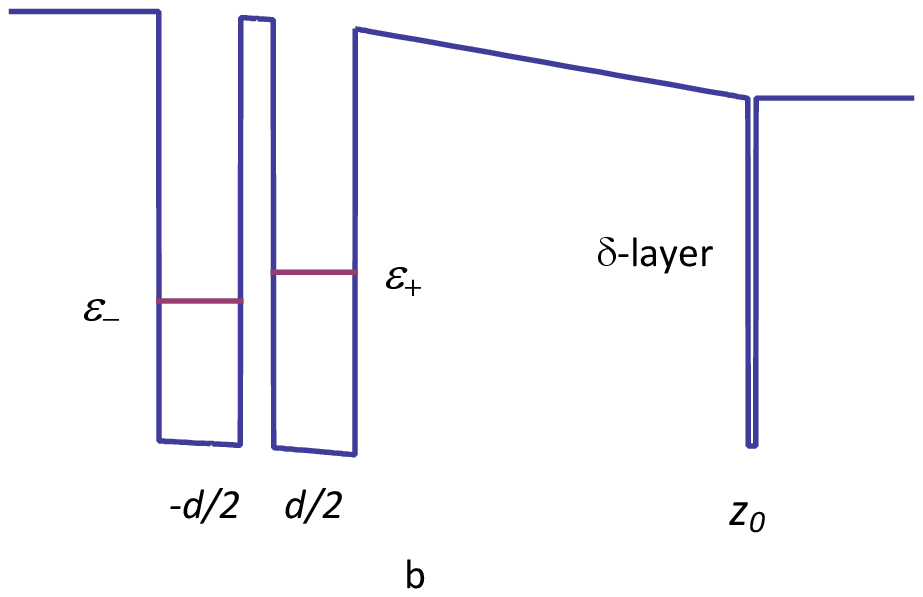}}
\centerline{\epsfxsize=5cm\epsfbox{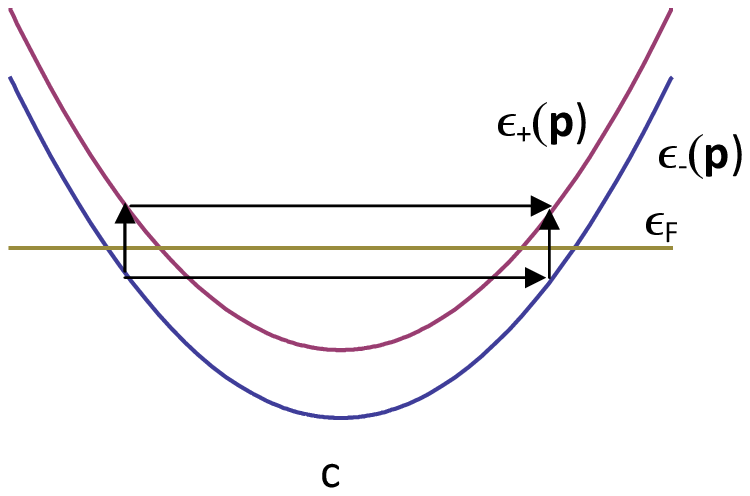}}
 \caption{(Color online) a) The sketch of the proposal experimental setup.  The electric field of light ${\bf E}(t)$ is tilted in (x,z) plane. The stationary current is directed along the x-axis. b) The sketch of the band structure. Quantum wells are centered in  planes $z=\pm d/2$. The carriers are provided by the $\delta$-layer of donors in plane $z_0$. c) The transition amplitude includes vertical transition caused by light between $\pm$ subbands and impurity scattering which does not conserve the in-plane momentum.}
\end{figure}
The paper is organized as follows. First, we will discuss a simple
classical model of the effect based on a parabolic well. Then, we
will find the transition probability in a classical electric
field. After that the current will be found using many-band
kinetic equation.
\section{Simple classical model}
To explain the physical origin of the effect
we consider a simple classical model instead of a 2D system: an electron in an oscillatory well in z-direction with confining potential $m\omega_0^2z^2/2$
affected by the alternating electric  field with x and z components
${\bf E}(t)=\mbox{Re}({\bf E}e^{-i\omega t})$.   The classical
Newton equation for an electron reads
\begin{equation}\label{New_eq}
 \ddot{\bf r}+\gamma\dot{\bf r}=e{\bf  E}/m,
\end{equation}
where we introduced the liquid friction coefficient
$\gamma=\gamma_0+\gamma_1z$. The dependence of the friction on $z$
takes into account the assumed weak asymmetry ($\gamma_0\gg\gamma_1z$)
of the well in $z$-direction.

 The forced solution of the Newton equation is found by expanding in  powers of $\gamma_1$:
$${\bf r}={\bf r}_0+{\bf r}_1+...,$$
$$z_0=\mbox{Re}\frac{eE_z}{m(-\omega^2+\omega_0^2-i\gamma_0\omega)}e^{-i\omega t}, $$
$$x_0=\mbox{Re}\frac{eE_x}{m(-\omega^2-i\gamma_0\omega)}e^{-i\omega t}, $$
\begin{equation}\label{x1}
\overline{(\dot{x}_1)}=\frac{\gamma_1\omega
e^2}{2\gamma_0m^2}\mbox{Im}\frac{E_x^*E_z}{(\omega^2-i\gamma_0\omega)(\omega^2-
\omega_0^2+i\gamma_0\omega)},~~~~~~\overline{(\dot{z}_1)}=0.\end{equation}
Here $\overline{(...)}$ denotes the time averaging.

Let  damping $\gamma_0$  be also small. Then the mean
velocity has a resonance at $\omega=\omega_0$. The frequency
behavior near this point depends on the kind of electromagnetic
field polarization: delta-like peak for the linear polarization  and
antisymmetric Fano-like resonance $\propto 1/(\omega-\omega_0)$
for circular polarization. The origin of this behavior is
explained by the character of the electron motion in the zero
approximation. Indeed, if $\gamma_1=0$, for linear polarization,
the electron rotates in the exact resonance  and vibrates along a
straight line out of resonance. For circular polarization the
behavior is opposite.

Liquid friction force $-\gamma\dot{\bf r}$ does not affect the
direction of vibrating motion; therefore it does not produce a
drift. At the same time, due to $\gamma_1,$ a rotating particle
differently brakes at the opposite (upper and lower) sides of the
circle that produces a translational displacement, and as a result, the mean drift. In the case of circular-polarized light, the
direction of the motion depends on the sign of polarization and
the sign of resonance detuning. The value of the drift velocity
near resonance does not depend on the friction strength, but it
depends on  ratio $\gamma_1/\gamma$.

Fig.2 illustrates this reasoning by the exact solution of the
Newton equation.
\begin{figure}[h]\label{fig2}
\centerline{\epsfxsize=3.5cm
\epsfbox{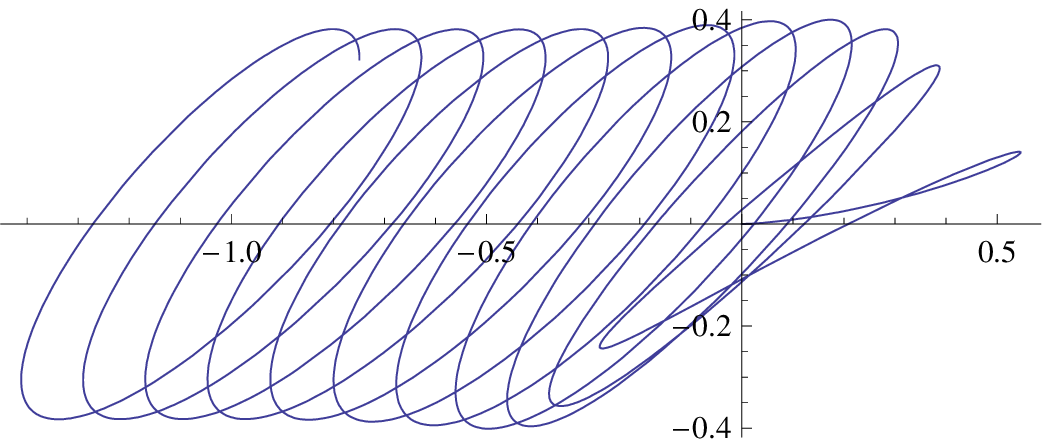}\epsfxsize=3.5cm\epsfbox{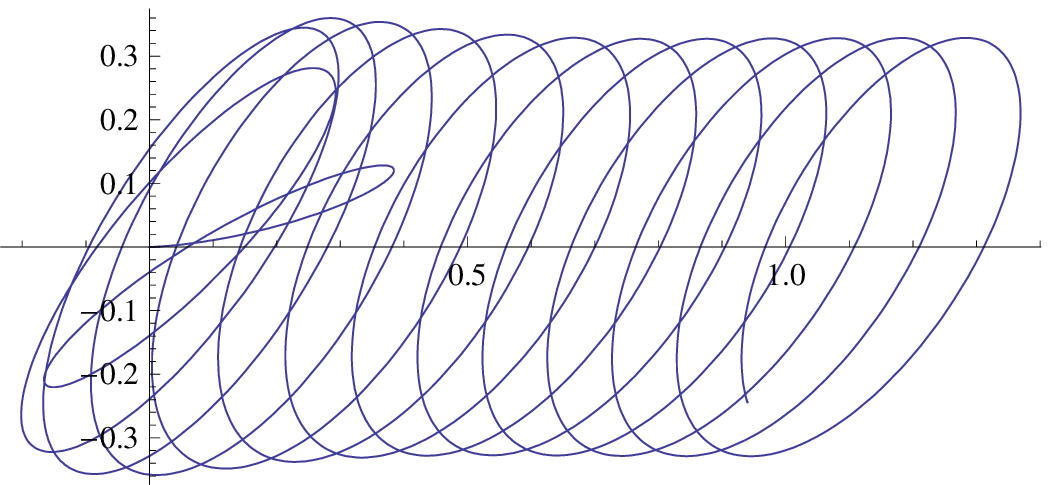}}
\centerline{\epsfxsize=3.3cm
\epsfbox{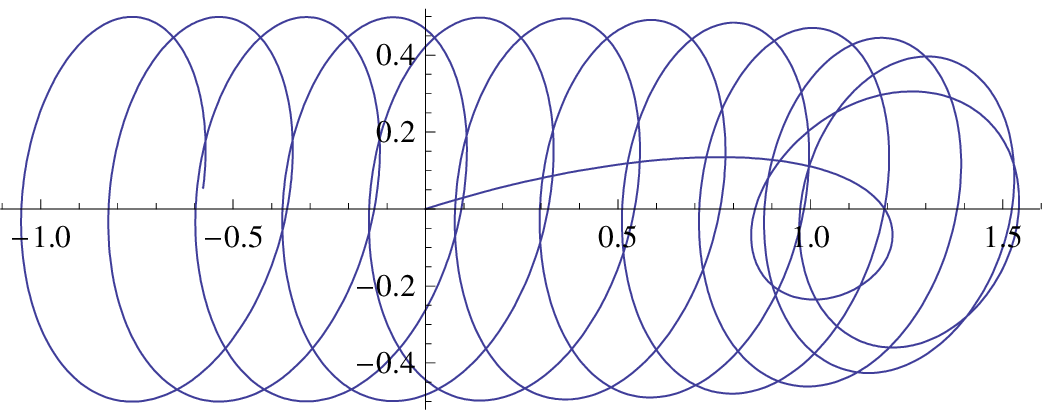}\epsfxsize=5.5cm\epsfbox{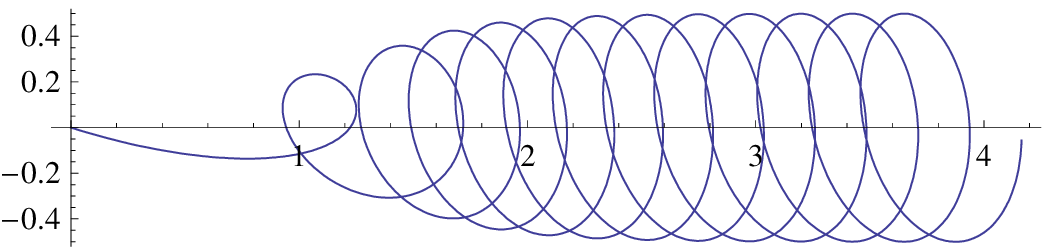}}
 \caption{ (Color online) Solution of the Newton equation for linear (c,d) and circular  (a,b) polarized electric field
 for initial conditions ${\bf r}(0)=0,~\dot{\bf r}(0)=0$ and parameters  $\gamma_0=0.2,~\gamma_1=0.02,~eE_x/m=0.3,~eE_z/m=0.1,~\omega_0=1$;
 for plot a)  $\omega=0.9$, for plot b) $\omega=1.1$, for plots a) and b) $\omega=1$.
For circular polarization the sign of detuning determines the
direction of the steady-state drift.}
\end{figure}
The  photogalvanic effect in this model has a purely classical
nature.  In particular, the circular PGE does not need the spin
pumping as in spin-related circular PGE. At the same time, the
classical and quantum photogalvanic effects  have different
properties.  The photogalvanic effect on intersubband transitions
of double quantum well Considered below has a resonant character
like the classical PGE discussed here. The difference is the
absence of  circular photogalvanic effect for transitions in a
double quantum well.

\section{Transitions between subbands of double well}
We study electrons with a parabolic isotropic energy spectrum in a
double quantum well (see Fig.1). The amplitude of transition
between wells is weak, but comparable to the separation of
energies of   individual wells.    The states with the in-plane
electron momentum ${\bf p}$ and the subband number $n=\pm$
$|n,{\bf p}> =\chi_n(z)\exp({i{\bf p}\ro})/\sqrt{S}$,  ($S$ is the
system area,  we set $\hbar=1$ throughout this section besides the
final expression) have energies $\epsilon_{n,\bf p}=p^2/2m +
\varepsilon_n$.  In this case, the subbands are parallel,
$\epsilon_{+,{\bf p}}-\epsilon_{-,{\bf p}}\equiv \varepsilon_+ -
\varepsilon_- $. This circumstance plays an important role in the
further consideration, providing the resonance of optical frequency
with a distance between subbands  for electrons with arbitrary
momenta. The overlapping of wave functions $\chi_n(z)$ is supposed
to be weak and  intersubband distance $\varepsilon_+ -
\varepsilon_-=\Delta$ ($\Delta>0$)  is small as compared to the
Fermi energy.  The scatterers (donors) are distributed in a delta-layer at
$z=z_0>0$. The well widths  and the distance $d$ between them are
assumed to be small as compared to $z_0$.

Assuming that the mean free time is large as compared to the distance between the levels of quantum wells (and also the Fermi energy) one can treat $n$ and ${\bf p}$ as good quantum numbers and describe the problem within the kinetic equation for distribution functions $f_{n,{\bf p}}$. In such an equation,  external classical alternating electric field ${\bf E}(t)=\mbox{Re}({\bf E}_0e^{-i\omega t})$ causes the transition between unpertubed states and determines the generation term in the kinetic equation. The interaction with charged impurities provides the  mechanism of electron scattering.
The kinetic equation reads
\begin{equation}\label{kin_eq}
    \sum_{n',{\bf p}'}W^{imp}_{n,{\bf p};n',{\bf p}'}(f^{(1)}_{n',{\bf p}'}-f^{(1)}_{n,{\bf
    p}})+ G_{n,{\bf p}} =0,
    \end{equation}
where the generation $G_{n,{\bf p}}$ is given by
\begin{equation} \label{G}
   G_{n,{\bf p}}= \sum_{n',{\bf p}'}W^{ph}_{n,{\bf p};n',{\bf p}'}(f^{(0)}_{n',{\bf p}'}-f^{(0)}_{n,{\bf
    p}}).
\end{equation}
Here $W^{imp}_{n,{\bf p};n',{\bf p}'}$ is the impurity transition
probability, $W^{ph}_{n,{\bf p};n',{\bf p}'}$ is the transition
probability due to the combined action of electromagnetic field and
impurities, $f^{(0)}_{n,{\bf p}}$ is the equilibrium distribution
function and $f^{(1)}_{n,{\bf p}}$ is the first correction to the
distribution function  in the external electromagnetic field. Using  the classical kinetic equation means  neglecting the
off-diagonal elements of the density matrix that is valid if the collision broadening of subbands
is less than the distance between them. The
perturbation includes the Hamiltonian of the interaction with
electromagnetic field $\hat{H}^{ph}$ and the potential energy of the electron interacting with impurities
$\hat{V}$. The first is
\begin{equation}\label{Hph}
    \hat{H}^{ph}=\frac{e}{c}\mbox{Re}\left({\bf A} e^{-i\omega t}\right)\hat{\bf v}\equiv \frac{1}{2}(\hat{U}e^{-i\omega t} + h.c.),
\end{equation}
where $\mbox{Re}({\bf A }e^{-i\omega t})$ is the vector potential
of electromagnetic field with frequency $\omega$, $\hat{\bf v}=(\hat{\bf v}^\|,\hat{v}^z)$ is
the velocity operator. The complex amplitude of electric field is
${\bf E}=i\omega {\bf A}/c$. Thus, the operator $\hat{U} = e({\bf
E\hat{v}})/i\omega$. Note that we suppose the electric field to be
homogeneous. The diagonal elements of in-plane components of the
the velocity operator $\bf v^\|_{n,{\bf p};n',{\bf p}'}={\bf
v}_{\bf p}\delta_{nn'}\delta_{{\bf p},{\bf p}'}$, ${\bf v}_{\bf p}=
\partial_{\bf p}\epsilon_{n,\bf p}= {\bf p}/m$.
 The normal component has
matrix elements $ v^z_{n,{\bf p};n',{\bf
p}'}=v^z_{n,n'}\delta_{{\bf p},{\bf p}'}$. The impurity
potential reads
\begin{equation}\label{V}
    V({\bf r})=\sum_iu({\bf r}-{\bf r}_i),
\end{equation}
where the sum runs over all the impurities situated in points
${\bf r}_i)$ with individual potentials $u({\bf r}-{\bf r}_i)$.

The appearance of the photogalvanic  current requires
non-conservation of the in-plane momentum in the electron
excitation process. Hence, the phototransitions should include
the participation of the ''third body''. In our case the
impurities play the role of this agent. The excitation probability
including the impurity scattering is determined by the
second-order transition amplitude. The needed term arises from the
interference of amplitudes caused by the $E_z$ and in-plane
components of the electric field. The draft of the transitions is
depicted in  Fig.1.

In the second order of the interaction,  the transition
probability is
\begin{eqnarray}\label{Wph}\nonumber
&&W^{ph}_{n,{\bf p};n',{\bf p}'}=\\\nonumber&&\frac{\pi}{2}\Big<\Bigg
|\sum_{n_1}\Bigg(\frac{V_{n,{\bf p};n_1,{\bf p}'}U^+_{n_1,{\bf
p};n',{\bf p}'}}{\eta +
i(\varepsilon_{n_1,n'}+\omega)}+ \frac{U^+_{n,{\bf
p};n_1,{\bf p}}V_{n_1,{\bf p};n',{\bf p}'}}{\eta +
i(\varepsilon_{n_1,n}-\omega)}\Bigg)
\Bigg|^2\Big>\times\\&& \delta(\epsilon_{n,{\bf p}}-\epsilon_{n',{\bf p}'}+\omega)
+ \nonumber \\&&\frac{\pi}{2} \Big<\Bigg |\sum_{n_1}\Bigg(\frac{V_{n,{\bf p};n_1,{\bf
p}'}U_{n_1,{\bf p};n',{\bf p}'}}{\eta +
i(\varepsilon_{n_1,n'}-\omega)}+ \frac{U_{n,{\bf
p};n_1,{\bf p}}V_{n_1,{\bf p};n',{\bf p}'}}{\eta +
i(\varepsilon_{n_1,n}+\omega)}\Bigg)
\Bigg|^2\Big>\times\nonumber\\&& \delta(\epsilon_{n,{\bf p}}-\epsilon_{n',{\bf
p}'}-\omega);   \ \ (\eta = +0).
\end{eqnarray}
Here $\varepsilon_{n_1,n}\equiv \varepsilon_{n_1}-\varepsilon_{n}$; angular brackets denote the average over  impurities configuration .
Using relations $U^+_{n,{\bf
p};n',{\bf p}'}=(U_{n',{\bf
p}';n,{\bf p}})^*, \ V_{n,{\bf
p};n',{\bf p}'}=(V_{n',{\bf
p}';n,{\bf p}})^*$  it is easy to prove that $W^{ph}_{n,{\bf p};n',{\bf p}'}=W^{ph}_{n',{\bf p}';n,{\bf p}}.$

The denominators in Eq.(\ref{Wph}) have their  resonance with the field frequency
independently from the  electron momentum. At the same time, the resonance
in the final state is absent due to non-conservation of the
in-plane momentum.

Eq.(\ref{Wph}) can be rewritten in the form (${\bf E}=({\bf E}_\|,E_z)$):
\begin{eqnarray}\label{Wph1}\nonumber
&&W^{ph}_{n,{\bf p};n',{\bf p}'}=\\&&\nonumber\frac{\pi e^2}{2\omega^2} \Big<\Bigg
|\sum_{n_1}\Bigg(V_{n,{\bf p};n_1,{\bf p}'}\Bigg(\frac{{\bf v}_{\bf p'}{\bf E}_\|^*\delta_{n_1,n'}}{i\omega}+\frac{v^z_{n_1,n'}E_z^*}{\eta +
i(\varepsilon_{n_1,n'}+\omega)}\Bigg)+ \nonumber \\\nonumber&& +\Bigg(\frac{{\bf v}_{\bf p}{\bf E}_\|^*\delta_{n,n_1}}{-i\omega}+\frac{v^z_{n,n_1}E_z^*}{\eta +
i(\varepsilon_{n_1,n}-\omega)}\Bigg)V_{n_1,{\bf p};n',{\bf p}'}\Bigg)
\Bigg|^2\Big>\times\\&&\nonumber\delta(\epsilon_{n,{\bf p}}-\epsilon_{n',{\bf p}'}+\omega)
+ \nonumber \\
&&\frac{\pi}{2} \Big<\Bigg |\sum_{n_1}\Bigg(V_{n,{\bf p};n_1,{\bf p}'}\Bigg(\frac{{\bf v}_{\bf p'}{\bf E}_\|\delta_{n_1,n'}}{-i\omega}+\frac{v^z_{n_1,n'}E_z}{\eta +
i(\varepsilon_{n_1,n'}-\omega)}\Bigg)+ \nonumber \\&& +\Bigg(\frac{{\bf v}_{\bf p}{\bf E}_\|\delta_{n,n_1}}{i\omega}+\frac{v^z_{n,n_1}E_z}{\eta +
i(\varepsilon_{n_1,n}+\omega)}\Bigg)V_{n_1,{\bf p};n',{\bf p}'}\Bigg)
\Bigg|^2\Big>\times\nonumber\\&&\delta(\epsilon_{n,{\bf p}}-\epsilon_{n',{\bf
p}'}-\omega).
\end{eqnarray}

It is evident that the contribution to photogalvanic effect is given by not the total transition probability $W^{ph}$ but only its odd in ${\bf p}, {\bf p}'$  part.
For  this part,   we have the following expression:
\begin{eqnarray}\label{Wpart}
    &&\tilde{W}^{ph}_{n,{\bf p};n',{\bf p}'}= \nonumber\\&&\nonumber \frac{\pi e^2}{\omega^3} \Bigg\{ \Big<\mbox{Re}\Bigg
[\sum_{n_1}\Bigg(V_{n,{\bf p};n',{\bf p}'}({\bf p}'-{\bf p}){\bf E}_\|^*\times\\&&\nonumber\Bigg(\frac{V^*_{n,{\bf p};n_1,{\bf p}'}v^z_{n',n_1}E_z}{i\eta +
(\varepsilon_{n_1,n'}+\omega)} + \frac{v^z_{n_1,n}E_z V^*_{n_1,{\bf p};n',{\bf p}'}}{i\eta +
(\varepsilon_{n_1,n}-\omega)}\Bigg)\Bigg)
\Bigg]\Big>\times\\&&\nonumber\delta(\epsilon_{n,{\bf p}}-\epsilon_{n',{\bf p}'}+\omega)
 \nonumber +\Big<\mbox{Re}\Bigg
[\sum_{n_1}\Bigg(V_{n,{\bf p};n',{\bf p}'}({\bf p}-{\bf p}'){\bf E}_\|\times\\&&\nonumber\Bigg(\frac{V_{n,{\bf p};n_1,{\bf p}'}^*v^z_{n',n_1}E_z^*}{i\eta +
(\varepsilon_{n_1,n'}-\omega)} + \frac{v^z_{n_1,n}E_z^* V_{n_1,{\bf p};n',{\bf p}'}^*}{i\eta +
(\varepsilon_{n_1,n}+\omega)}\Bigg)\Bigg)
\Bigg]\Big>\times\\&&\delta(\epsilon_{n,{\bf p}}-\epsilon_{n',{\bf p}'}-\omega)\Bigg\}.
\end{eqnarray}

Kinetic equation Eq.(\ref{kin_eq}) can be transformed to
\begin{equation}\label{kin_eq1}
   \frac{1}{ \tau_n(p)}f^{(1)}_{n,{\bf p}}-\frac{1}{\tau_{n,-n}(p)}f^{(1)}_{-n,{\bf p}}=G_{n,{\bf p}},
    \end{equation}
    where $\tau_n(p)$ is the intra-subband transport relaxation time and $\tau_{n,-n}(p)$ is the time of transition  from the state $(n,{\bf p})$ to all states of the subband $(-n)$. These values are determined by
    \begin{eqnarray}\label{nuintra}\nonumber
    \frac{1}{ \tau_n(p)}= 2\pi \sum_{\bf p'}\Big[\Big<|V_{n,{\bf p};n,{\bf p}'}|^2\Big>\delta(\epsilon_{n,{\bf p}}-\epsilon_{n,{\bf p}'})(1-\frac{{\bf p} {\bf p}'}{p^2})\\\nonumber
      + \Big<|V_{n,{\bf p};-n,{\bf p}'}|^2\Big>\delta(\epsilon_{n,{\bf p}}-\epsilon_{-n,{\bf p}'})\Big];\\
        \label{nuinter} \frac{1}{\tau_{n,-n}(p)} = 2\pi \sum_{\bf p'}\Big<|V_{n,{\bf p};-n,{\bf p}'}|^2\Big>\delta(\epsilon_{n,{\bf p}}-\epsilon_{-n,{\bf p}'})\frac{{\bf p} {\bf p}'}{p^2}.
    \end{eqnarray}

    Solving Eq.(\ref{kin_eq1}) we find (argument ${\bf p}$ is omitted):
    \begin{equation} \label{f+-}
        f^{(1)}_n =\Big(G_n \tau_n + G_{-n}\frac{\tau_+\tau_-}{\tau_{n,-n}}\Big)\Big(1-\frac{\tau_+\tau_-}{\tau_{+,-}\tau_{-,+}}\Big)^{-1}.
    \end{equation}

    The expressions for $\tau_n(p),  \tau_{n,-n}(p)$ and $\tilde{W}^{ph}_{n,{\bf p};n',{\bf p}'}$ contain correlators  of the form
         $ \Big<V_{n,{\bf p};n',{\bf p}'} V_{m,{\bf p};m',{\bf p}'}\Big>$. In the case of impurities situated in layer $z=z_0 \ ({\bf r}_i=({\ro}_i,z_0))$ the function $V({\bf r})$ reads
         \begin{equation}\label{V1}
            V({\bf r})=\sum_{{\bf q},i}u_{\bf q}e^{-q|z-z_0|}\exp{(-i{\bf q}({\ro }-{\ro_i}))},
         \end{equation}
         where $u_{\bf q}$ is the 2D Fourier component of  the impurity center potential. For example, for unscreened Coulomb center $u_{\bf q}=2\pi e^2/\kappa  qS$ ($\kappa $ is the background dielectric constant).
         Correlators are given by
         \begin{eqnarray}\label{corr}
               &\Big<V_{n,{\bf p};n',{\bf p}'} V^*_{m,{\bf p};m',{\bf p}'}\Big>= n_sS\int dz dz' |u_{{\bf p}-{\bf p}'}|^2e^{-q(2z_0-z-z')}\nonumber\\&\times \chi_n(z)\chi_{n'}(z)\chi_m(z')\chi_{m'}(z').
    \end{eqnarray}
Here  $n_s$ is the  areal density of scatterers.
       We suppose that the electron wavelength is larger than $d$. In this approximation one can  find from Eq.(\ref{corr}):
      \begin{eqnarray}\label{corr1}
        \Big<V_{n,{\bf p};n',{\bf p}'} V^*_{m,{\bf p};m',{\bf p}'}\Big>= n_sS|u_{{\bf p}-{\bf p}'}|^2e^{-2qz_0}\times\nonumber\\\Big[\delta_{n,n'}\delta_{m,m'}+  q(z_{n,n'}\delta_{m,m'}+z_{m,m'}\delta_{n,n'})\Big].
      \end{eqnarray}
        Matrix elements $z_{nn'}$ should be estimated for  specific wave functions. For simplicity, we will use the wave functions of two delta-functional wells in the tight-binding approximation. The seed  states with energies $\varepsilon_0\pm \Delta_0/2$  can be written as
       \begin{equation}\label{basis}\chi_{1,2}=\sqrt{\kappa}e^{-\kappa|z\mp d/2|}.
       \end{equation}
       In basis (\ref{basis}) $\chi_+=(1,\beta)/\sqrt{1+\beta^2}$, $\chi_-=(\beta,-1)/\sqrt{1+\beta^2}$, where  $\beta$ is the mixing amplitude.  The corresponding states energies  are $\varepsilon_\pm=\varepsilon_0\pm\Delta/2$,  $\Delta=\sqrt{\Delta_0^2+4t_0^2}$, where $t_0\sim\varepsilon_0 e^{-\kappa d}$ is a  hopping amplitude between wells. For  quantity $\beta$ we have $\beta=2t_0/(\Delta+\Delta_0)$. The matrix elements of $z$ are $z_{++}=-z_{--}=d(1-\beta^2)/(2(1+\beta^2))$, $z_{+-}=\beta d/(1+\beta^2)$.

         Inserting Eq.(\ref{corr1}) in Eq.(\ref{nuintra}) we get the          expressions for $\tau_+ \approx \tau_-=\tau$ and a small difference $1/\tau_--1/\tau_+$:
         \begin{eqnarray}\label{tau}
        &&\frac{1}{\tau}= m n_s\int\frac{d{\bf q}}{2\pi}|\tilde{u}_{\bf q}|^2e^{-2qz_0}\delta(q^2+2{\bf pq})\frac{q^2}{p^2}\nonumber \\
                 && \frac{1}{\tau_-}-\frac{1}{\tau_+}\nonumber= m(z_{++}-z_{--}) n_s\nonumber\times\\ &&\int\frac{d{\bf q}}{\pi}|\tilde{u}_{\bf q}|^2e^{-2qz_0}\delta(q^2+2{\bf pq})\frac{q^3}{p^2},
         \end{eqnarray}
         where $\tilde{u}_{\bf q}=Su_{\bf q}$.
       From Eq.(\ref{corr1}) it is seen that $\tau_{n,-n}\gg \tau_n$ and, so, Eq.(\ref{f+-}) can be simplified
      \begin{equation}\label{f+-1}
        f^{(1)}_n = G_n \tau_n .
      \end{equation}
      Further we will consider the resonance situation when frequency $\omega$ is close to $\Delta$.  Smallness $\Delta,$
 as compared to the Fermi energy $\epsilon_F=p_F^2/2m \ (p_F$ being the Fermi momentum) leads to approximate expressions for $G_+\approx -G_-$,
 \begin{eqnarray}\label{G1}
    G_+= \frac{n_se^2\Delta z_{+-}^2}{\pi  \omega^2}\frac{\eta}{(\Delta-\omega)^2+\eta^2}\frac{\partial f_{\bf p}^{(0)}}{\partial \mu}\times\nonumber\\\int d{\bf q}({\bf q}\cdot \mbox{Re}({\bf E}_\|E_z^*))q|\tilde{u}_{\bf q}|^2e^{-2qz_0}\delta(q^2+2{\bf pq}).
 \end{eqnarray}
As a result, for the current of photogalvanic effect, we have
\begin{eqnarray}\label{j1}
  {\bf j}= -\frac{n_se^3\Delta z_{+-}^2}{4\pi^3 m \omega^2}\frac{(\tau_+-\tau_-)\tau}{(\Delta-\omega)^2\tau^2+1} \mbox{Re}({\bf E}_\|E_z^*)\times \nonumber\\\int d{\bf p}\frac{\partial f_{\bf p}^{(0)}}{\partial \mu} \int d{\bf q}|\tilde{u}_{\bf q}|^2e^{-2qz_0}q^3 \delta(q^2+2{\bf qp} ).
 \end{eqnarray}
 Eq. (\ref{j1}) has a resonant character with the resonance at $\omega=\Delta$. This resonance results from the intermediate state for transition due to the parallelism (equidistance) of subbands. The resonance  is smeared due to scattering, e.g., by impurities. To include this smearing, the infinitesimal $\eta$ was replaced by finite relaxation rate $1/\tau$ which can be estimated from mobility. This leads to the finiteness of the current at the point of resonance.

 At temperature $T=0$ the latter expression is simplified, and we obtain the final result for the required value:
 \begin{equation}\label{j_fin}
    {\bf j}= -\frac{4e^3(z_{++}-z_{--})\Delta z_{+-}^2\epsilon_F\tau}{\pi \omega^2d^2((\Delta-\omega)^2\tau^2+1)}\mbox{Re}({\bf E}_\|E_z^*)F,
 \end{equation}
 where we  introduced a dimensionless quantity $F=d^2\Phi_3^2\Phi_2^{-2}$,
 \begin{equation}\label{Phi}
\Phi_s=\int_0^{2p_{F}} dq q^s|\tilde{u}_{\bf q}|^2e^{-2qz_0}\frac{1}{\sqrt{1-q^2/4p_F^2}}.
 \end{equation}
 In the specific case of $p_Fz_0\gg 1$ Eq. (\ref{Phi}) is reduced to
\begin{equation}\label{Ph1}
  \Phi_s=\int_0^{\infty} dq q^s|\tilde{u}_{\bf q}|^2e^{-2qz_0}.
 \end{equation}  If the scattering is determined by the charged non-screened impurities  $ F=d^2/4z_0^2. $

In the model of two $\delta$-like wells Eq.(\ref{j_fin}) gives For linear polarized wave
\begin{equation}\label{delt_fin}
 {\bf j}= -\frac{4e^3d \beta(1-\beta^2 )n}{m(1+\beta^2)^2 \Delta}\frac{\tau}{(\Delta/\hbar-\omega)^2\tau^2+1} E_0^2\sin{(2\theta)}F,
\end{equation}
where $E_0$ is the amplitude of the electric field, $\theta$ is the angle between the field and the normal to the system.

 It should be emphasized that the current contains the linear response only.
This  distinguishes the quantum double-well result from the simple classical model of the effect considered in the previous section.

Let us estimate the value of the effect. Considering $\beta$ as a free parameter we can choose $\beta=\sqrt{2}-1$ to maximize the $\beta$-dependent factor in Eq.(\ref{delt_fin}) $\beta(1-\beta^2)/(1+\beta^2)^2=1/4$.  The optimum for PGE observation corresponds to $\omega=\Delta$ and $\theta=\pi/4$. Choosing the  typical values for $GaAs/AlGaAs$ double quantum wells $d=5\cdot 10^{-7}$ cm, $z_0=3\cdot 10^{-6}$cm,  $\epsilon_F$=20 meV ($n=6.2\cdot 10^{11}$cm${}^{-2}$), $\Delta$=0.1 meV,  $\tau=4\cdot 10^{-11}$ s, $E_0=1$ V/cm we find for this optimal situation $j\approx 3.6~\mu$A/cm that is a quite measurable value.

It should be emphasized that  the initial and final states in the transition  can belong to the different or the same subbands. The resonant behavior results from the resonance on the
intermediate state rather than the energy conservation in the
final states, because the conservation law for the phototransition with the participation of impurity scattering does not give a fixed frequency for the transition. The sharpness of the resonance is conditioned  by
the equidistance of the energy bands in a 2D well.
\section{Conclusions }
We  found the stationary current along a double-well system
affected by the linear-polarized far-infrared wave. The stationary current
originates from the periodic vibration of electrons between two
non-equivalent quantum wells caused by the normal component of the
alternating electric field with synchronic in-plane
acceleration/deceleration by the in-plane component of the electric field.
The linear photogalvanic effect needs vertical asymmetry of
the quantum well.  The effect has the peak resonant structure
 connected with the parallel subbands of the double quantum well.
 The resonant frequency can be easily tuned by the application of the gate voltage.
 The optimal range of frequencies is $10^{11}\div$ $10^{13}\div$ s${}^{-1}$.
 The predicted value of the current is experimentally measurable.

\subsection*{Acknowledgements}
This research was supported  by RFBR grant nos. 11-02-00060, 11-02-00730 and 11-02-12142.


\begin{thebibliography}{30}
\bibitem{malin} V.I.Belinicher, V.K.Malinovskii, and B.I.Sturman, Sov. Phys.
JETP {\bf 46}, 362 (1977).
\bibitem{we} E.M.Baskin, M.D.Blokh, M.V.Entin, and L.I.Magarill, Phys.
Status Solidi B {\bf 83}, K97 (1977).
\bibitem{we1} E.M.Baskin, L.I.Magarill, and M.V.Entin, Sov. Phys. Solid
State {\bf 20}, 1403 (1978).
\bibitem{belin} V.I.Belinicher and B.I.Sturman, Sov.Phys.Usp. {\bf 23}, 199
(1980) [Usp.Fiz.Nauk {\bf 130}, 415 (1980).
\bibitem{ivch} E.L.Ivchenko and G.E.Pikus, in Semiconductor Physics, edited
by V. M. Tushkevich and V. Ya. Frenkel (Cons. Bureau, New
York, 1986), p. 427.
\bibitem{sturm} B.I. Sturman and V.M. Fridkin, The Photovoltaic and
Photorefractive Effects in Non-Centrosymmetric Materials
(Nauka, Moscow, 1992; Gordon and Breach, New
York, 1992).
\bibitem{ivch1} E.L. Ivchenko, Optical spectroscopy of semiconductor
nanostructures (Alpha Science International, Harrow,
UK, 2005).
\bibitem{ivch2} E.L.Ivchenko and G.E.Pikus, in Superlattices and Other Heterostructures, Springer Series in Solid-State Sciences, Vol. 110
(Springer, Berlin, 1997).
\bibitem{gan} S.D.Ganichev and W.Prettl, Intense Terahertz Excitation
of Semiconductors (Oxford University Press, 2006).

\bibitem{shepel} A.D.Chepelianskii, and D.L.Shepelyansky, Phys.Rev.B {\bf 71}, 052508 (2005).
\bibitem{shepel1} G.Cristadoro, D.L.Shepelyansky, Phys.Rev.E {\bf 71},
036111 (2005).
\bibitem{we2}    M.V.Entin and L.I.Magarill, Phys.Rev.B {\bf 73}, 205206 (2006).
\bibitem{shepel2} A.D.Chepelianskii, M.V.Entin, L.I.Magarill, and D.L.Shepelyansky, Phys.Rev.E {\bf 78}, 041127 (2008).
 \bibitem{sassin} S.Sassine, Yu.Krupko, J.-C.Portal, Z.D.Kvon, R.Murali,
K.P.Martin, G.Hill and A.D.Wieck, Phys.Rev.B {\bf 78}, 045431 (2008).
\bibitem{ivch3} H.Diehl, V.A.Shalygin, L.E.Golub, S.A.Tarasenko, S.N.Danilov, V.V.Bel'kov, E.G.Novik, H.Buhmann,
C.Brune, L.W.Molenkamp, E.L.Ivchenko, and S.D.Ganichev, Phys.Rev.B {\bf 80}, 075311 (2009).
\bibitem{taras} P.Olbrich, S.A.Tarasenko, C.Reitmaier, J.Karch, D.Plohmann, Z.D.Kvon, and S.D.Ganichev, Phys.Rev.B {\bf 79}, 121302(R) (2009).
\bibitem{we3}    L.E.Golub, S.A.Tarasenko, M.V.Entin, and L.I.Magarill,
Phys.Rev.B {\bf 84}, 195408 (2011).
\bibitem{karch} J. Karch, S.A.Tarasenko, E.L.Ivchenko, J.Kamann, P.Olbrich, M.Utz, Z.D.Kvon, and S.D.Ganichev, Phys.Rev.B {\bf 83}, 121312(R) (2011)
\bibitem{we4} L.I.Magarill and M.V.Entin, Fiz. Tverd. Tela (Leningrad)
{\bf 21}, 1280 (1979) [Sov. Phys. Sol. State {\bf 21}, 743 (1979)].
\bibitem{we5} L.I.Magarill and M.V.Entin, Poverkhnost'. Fizika, khimiya, mekhanika, {\bf 1}, 74 (1982).
\bibitem{taras1} S.A.Tarasenko, Phys.Rev. B {\bf 83}, 035313 (2011)
\bibitem{alper} V.L.Al'perovich. V.I.Belinicher, V.N.Novikov, and A.S.Terekhov, Zh. Eksp. Teor. Fiz. {\bf 80}, 2298 (1981) [Sov. Phys.
JETP {\bf 53}, 1201 (1981)].
\bibitem{gus} G.M.Gusev, Z.D.Kvon, L.I.Magarill, A.M.Palkin, V.I.Sozinov, O.A.Shegai, and V.M.Entin, JETP Lett. {\bf 46}, 33
(1987).
\bibitem{we6} L.I.Magarill, V.M.Entin, Sov. Phys. Solid State {\bf 31}, 1299 (1989).
\end{thebibliography}
\end{document}